\documentclass[journal,twoside,web]{ieeecolor}
\usepackage{generic}
\usepackage{cite}

\usepackage{amsmath,amssymb,amsfonts}
\newcommand{\var}[1]{{\operatorname{#1}}}
\usepackage{mathptmx}

\usepackage{graphicx}
\usepackage{verbatim}
\usepackage[caption=false,font=footnotesize,subrefformat=parens,labelformat=parens]{subfig}
\usepackage{multirow}
\usepackage{booktabs}
\usepackage{threeparttable}
\usepackage{diffcoeff}
\usepackage{textcomp}
\usepackage{siunitx}
\usepackage{color,soul}
\usepackage[capitalize]{cleveref}

\def\BibTeX{{\rm B\kern-.05em{\sc i\kern-.025em b}\kern-.08em
    T\kern-.1667em\lower.7ex\hbox{E}\kern-.125emX}}
    
\begin{document}
\title{A Memory Window Expression to Predict the Scaling Trends and Endurance of FeFETs}
\author{Nicolò Zagni, \IEEEmembership{Student Member, IEEE}, Paolo Pavan, \IEEEmembership{Senior Member, IEEE}, and Muhammad Ashraful Alam, \IEEEmembership{Fellow, IEEE}
\thanks{N. Zagni was visiting scholar at the School of Electrical and Computer Engineering, Purdue University, West Lafayette, IN 47907 USA. He is with the Department of Engineering “Enzo Ferrari”, University of Modena and Reggio Emilia, via P. Vivarelli 10, 41125 Modena (MO), Italy. (e-mail: nicolo.zagni@unimore.it).}
\thanks{P. Pavan is with the Department of Engineering “Enzo Ferrari”, University of Modena and Reggio Emilia, via P. Vivarelli 10, 41125 Modena (MO), Italy.}
\thanks{M. A. Alam is with the School of Electrical and Computer Engineering, Purdue University, West Lafayette, IN 47907 USA.}}
\maketitle

\begin{abstract}
    The commercialization of non-volatile memories based on ferroelectric transistors (FeFETs) has remained elusive due to scaling, retention, and endurance issues. Thus, it is important to develop accurate characterization tools to quantify the scaling and reliability limits of FeFETs. In this work, we propose to exploit \textcolor{black}{an} analytical expression for the Memory Window (\emph{MW}, i.e., the difference between the threshold voltages due to polarization switching) as a tool to: \emph{i)} identify a universal scaling behavior of \emph{MW} regardless of the ferroelectric material; \emph{ii)} give an alternative explanation for \emph{MW} being lower than theoretical limits; based on it, \emph{iii)} devise strategies to maximize \emph{MW} for a given ferroelectric thickness; and \emph{iv)} predict endurance and explain its weak dependence on writing conditions (under specific assumptions). According to these findings, the characterization and analysis of \emph{MW} would enable the systematic comparison and development of next-generation FeFET based on emerging ferroelectric materials.
\end{abstract}

\section{Introduction}\label{sec:introduction}
\IEEEPARstart{T}{he} first demonstration of a thin-film ferroelectric transistor (FeFET) by Moll and Tarui in the early 1960s fueled the tantalizing promise of Non-Volatile Memories (NVMs) based on such technology \cite{Moll1963}. Successive generations of FeFETs have revived the interest of the community \cite{Okuyama2016,Dunkel2017,Ali2018}, only to realize that the technological issues related to scaling, retention, and endurance hindered commercialization. Although FeFETs offer better nonvolatility, scaling potential, higher read-write speeds and lower dissipation power over memory devices such as DRAM, SRAM and Flash memory \cite{Okuyama2016}, their reduced retention and endurance compared to other novel NVMs such as Resistive RAMs and Phase-Change RAMs \cite{Salahuddin2018} have restricted their adoption. Nonetheless, the successful demonstration of a CMOS-compatible FeFET would advance a broad range of applications, such as: \emph{i)} Logic-In-Memory (LiM) circuits \cite{Yin2016}; \emph{ii)} artificial neural networks \cite{Jerry2018,Aziz2018}; and \emph{iii)} Ternary Content Addressable Memories (TCAMs) \cite{Yin2016,Aziz2018}. Given the 50+ years history and the potential of the technology, it is important to develop accurate characterization tools to quantify endurance/retention and identify the limits of latest generation FeFETs based on HfO\textsubscript{2} and ZrO\textsubscript{2} binary oxides \cite{Muller2012}. 

In this work, we derive an analytical expression for the $MW$ that can be used to interpret experiments, quantify the scaling limits of FeFETs as well as the endurance of such devices. The $MW$ is a useful metric that allows comparing the performance of FeFETs regardless of the application or the specific technology. The theoretical framework for ferroelectric MOSFETs employed in this work is based on the Landau-Devonshire theory, which was originally developed to explain the operation of Negative Capacitance transistors (NCFETs) \cite{Salahuddin2008} and later exploited to model the operation of hysteretic ferroelectric transistors with a simplified structure \cite{Chen2011}. Here, we generalize the approach presented in \cite{Chen2011} to realistic Metal-Ferroelectric-Insulator-Semiconductor (MFIS) stacks to derive a simple expression for $MW$ that under specific assumptions and approximations can be used as a tool to: \emph{i)} identify a universal scaling behavior regardless of the ferroelectric material; \emph{iii)} explain why the \emph{MW} is lower than predicted theoretical values; based on it, \emph{iv)} devise strategies to maximize \emph{MW} for a given ferroelectric thickness; and \emph{iv)} predict endurance and explain its weak dependence on writing conditions. The analysis provides insights into the features of FeFET that generally are left unveiled by results based on TCAD simulations. 

The paper is organized as follows. In \cref{sec:anal_model} we discuss the derivation of the analytical model, the limits of validity of the approach, and the design guidelines to maximize $MW$. In \cref{sec:results} we present the results in terms of $MW$ scaling and endurance. In \cref{sec:conclusions}, we draw the conclusions of the work. In the Appendix we show the validation of the analytical expressions with numerical simulations and the comparison with the Preisach model for ferroelecrics.
\section{Analytical Model}\label{sec:anal_model}
\subsection{Derivation of V\textsubscript{th,on}, V\textsubscript{th,off} and MW Expressions}\label{sec:anal_derivation}
In this section, we present the derivation of the on- and off- threshold voltage, $V_{th,on}$, $V_{th,off}$, and $MW$ analytical expressions for the Metal-Ferroelectric-Insulator-Semiconductor (MFIS) stack, which is the most common device structure for FeFETs \cite{Dunkel2017} (another option includes an additional metal layer between the ferroelectric and the insulator (MFMIS) \cite{Salahuddin2008,Pahwa2018,Zagni2019}). From a physical point of view, $MW$ is determined by the polarization switching of the ferroelectric layer present in the gate stack and the state of the memory is encoded as the channel conductance at a particular gate bias, i.e., $V_{READ}$ (for Non-Volatile Memories, $V_{READ}\approx0$). In this context, the $MW$ is defined simply as $MW\equiv{}V_{th,on}-V_{th,off}$. The $MW$ is derived by generalizing the approach followed in \cite{Chen2011}, with the inclusion of \emph{i)} the SiO\textsubscript{2} interface layer between the ferroelectric layer and the semiconductor channel, and \emph{ii)} the linear component of the ferroelectric layer \cite{Hoffmann2018}. The analytical $MW$ expression allows identifying the key parameters that influence the scaling trends of FeFETs, as explained later.

The derivation starts with the model of the electrostatic behavior of the FeFET, obtained by coupling the classical MOSFET surface potential equation (SPE) with the Landau-Devonshire theory \cite{Chen2011,Pahwa2017}:
\begin{equation}\label{eq:spe}
    V_{GS}-V_{FB}=V_{ins}+\psi_s
\end{equation}
where $V_{GS}$ is the applied gate bias, $V_{FB}$ is the flatband voltage, $V_{ins}$ is the insulator voltage (including both ferroelectric and oxide interface layer), and $\psi_s$ is the surface potential. $V_{ins}$ is expressed as follows:
\begin{equation}\label{eq:vins}
    V_{ins}=Q_s\left(\frac{1}{C_{LD}+C_{FE}}+\frac{1}{C_{ox}}\right)
\end{equation}
where $C_{ox}=\varepsilon_{ox}/t_{ox}$ is the oxide interface layer capacitance and
\begin{equation}\label{eq:fe_cap}
    C_{LD}=\frac{1}{t_{FE}(2\alpha+12\beta{}Q_s^2)}\quad{}C_{FE}=\frac{\varepsilon_{FE}}{t_{FE}}
\end{equation}
are the capacitance components of the ferroelectric due to polarization (obtained from Landau-Devonshire theory), $C_{LD}$, and linear dielectric behavior \cite{Hoffmann2018}, $C_{FE}$, respectively. $Q_s$ is the semiconductor charge, $\alpha$, $\beta$ are the Landau parameters for the ferroelectric layer, $\varepsilon_{FE}$ is the linear dielectric constant of the ferroelectric layer, and $t_{FE}$ is the ferroelectric thickness.

To reach closed-form expressions for $V_{th,on}$, $V_{th,off}$ and $MW$, we simplify the $Q_s$ expression by considering only the inversion layer charge \cite{Chen2011}. The on-threshold voltage, $V_{th,on}$, is obtained by solving $\diffp{V_G}/{\psi_s}=0$, which is the condition at which the FeFET is about to enter the so-called negative capacitance region \cite{Salahuddin2008}. Because the total gate capacitance is negative, and therefore unstable, the ferroelectric switches to the saturated polarization value (skipping the negative capacitance region), turning on the device: this is the $V_{th,on}$ condition. The final expression is then written as [neglecting higher-order terms in \eqref{eq:vins} and \eqref{eq:fe_cap}]: 
\begin{equation}\label{eq:vth_on}
    V_{th,on}=V_{FB}+2V_{t}\ln{\left(\frac{2V_{t}}{|a|Q_0}\right)}-2V_{t}
\end{equation}
where $V_{t}=k_{B}T\text{/}{q}$ is the thermal voltage, $k_B$ is the Boltzmann constant, $T$ is the device temperature, $q$ is the elementary charge, and $a\equiv2\alpha{}t_{FE}\text{/}(1+2\alpha\varepsilon_{FE})+1\text{/}C_{ox}$. $Q_0=\sqrt{2\varepsilon_{s}k_{B}Tn_i^2\text{/}{N_a}}$ is the pre-exponential term of the inversion charge expression, namely $Q_s=Q_0\exp{\left(\psi_s\text{/}{2V_{t}}\right)}$, where $\varepsilon_s$ is the semiconductor dielectric constant, $n_i$ is the intrinsic carrier density, and $N_a$ is the substrate doping density (we consider a p-type substrate for a NMOS device). The off-threshold voltage, $V_{th,off}$, is obtained instead by solving $\diffp{Q_s}/{V_{ins}}=0$ (with $Q_s>0$). At this condition the FeFET is again at the boundary of the negative capacitance region, but in the opposite direction with respect to the previous case (for $V_{th,on}$), and the ferroelectric switching causes the device to turn off. (A more detailed discussion on why the condition for on- and off- switching are non-symmetrical is found in \cref{sec:anal_non_symmetric}). The final expression is as follows:
\begin{equation}\label{eq:vth_off}
    V_{th,off}=V_{FB}+2V_{t}\ln{\left(\frac{Q_{sw}}{Q_0}\right)}-V_{sw}
\end{equation}
where $V_{sw}$ is the switching voltage, defining the boundary between the positive and negative capacitance region, occurring at the switching charge $Q_{sw}$.
Finally, the memory window expression, $MW$, is obtained simply by subtracting \eqref{eq:vth_off} from \eqref{eq:vth_on}:
\begin{equation}\label{eq:mw}
    MW=2V_{t}\ln{\left(\frac{2V_t}{|a|Q_{sw}}\right)}+\left(V_{sw}-2V_t\right).
\end{equation}
This expression is the key result of the paper, determined by the $\psi_s$ and the $V_{ins}$ difference between the on- and off- switching conditions, corresponding to the first and second term in \eqref{eq:mw}, respectively. Interestingly, the $MW$ does not depend on $V_{FB}$, nor on $N_a$ because they both affect $V_{th,on}$ and $V_{th,off}$ equally. As revealed by \eqref{eq:mw}, $MW$ primarily depends on $V_{sw}$ which gives rise to a universal scaling trend for the $MW$. Approximate expressions for $V_{sw}$, $Q_{sw}$ that connect them to the ferroelectric parameters can be derived by considering $C_{LD}\gg{}C_{FE}$:
\begin{subequations}\label{eq:vsw}
    \begin{align}
        V_{sw}\equiv{}&-\left(aQ_{sw}+b\,Q_{sw}^3\right)=\frac{2}{3}|\,a|Q_{sw}\\
        a\equiv{}2\alpha{}t_{FE}+&\frac{1}{C_{ox}}\quad{}b\equiv4\beta{}t_{FE}\quad{}Q_{sw}\equiv{}\sqrt{\frac{|a|}{3b}}.
    \end{align}    
\end{subequations}
\subsection{Design Constraints to Guarantee MW $>0$}\label{sec:anal_constraint}
\begin{figure}[t!]
    \centering{
        \includegraphics[width=\columnwidth]{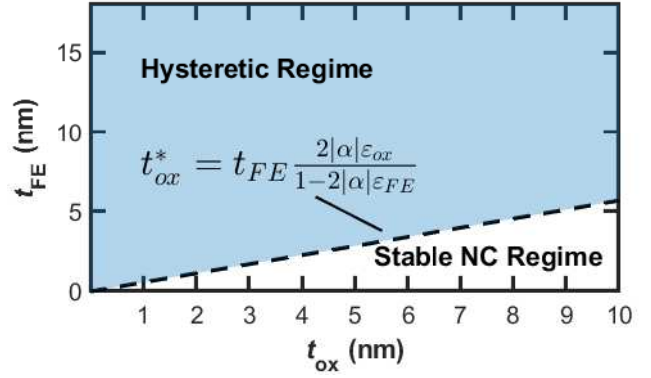}
    }
    \caption{Design space of FeFET based on the criterion imposed by \eqref{eq:tox_ineq} to guarantee $MW>0$. If for a given $t_{FE}$, $t_{ox}<t_{ox}^*$ (black dashed line) then the device stabilizes in the hysteretic regime and memory operation can be achieved. Conversely, the theory predicts that stable negative capacitance (NC) regime can occur if the opposite condition is satisfied. Parameters are given in \cref{tab:sim_param}.}
    \label{fig:1}
\end{figure}
    As mentioned in \cref{sec:introduction}, the Landau-Devonshire formalism is conventionally adopted to describe the behavior of NCFETs \cite{Salahuddin2008, Pahwa2017, Hoffmann2018} which occurs under particular conditions leading to $\psi_s$ amplification (for a given $V_{GS}$) and sub-threshold swing, $SS$, below the Boltzmann limit of \SI{60}{\milli\volt/dec}. The theory, however, allows describing also the hysteretic behavior of ferroelectrics that occurs when the total gate capacitance is negative \cite{Alam2019,Rusu2016}. Approximately, FeFET memory operation is guaranteed by the following inequality:
    \begin{equation}\label{eq:hyst_cond}
        MW>0\Leftrightarrow\left(\frac{1}{C_{LD}+C_{FE}}+\frac{1}{C_{ox}}\right)<0
    \end{equation}
    which can be more conveniently expressed in terms of $t_{FE}$ and $t_{ox}$ as follows (by neglecting higher-order terms):
    \begin{equation}\label{eq:tox_ineq}
        t_{ox}<t_{ox}^*=t_{FE}\frac{2|\alpha|\varepsilon_{ox}}{1-2|\alpha|\varepsilon_{FE}}
    \end{equation}
    The constraint defined by \eqref{eq:tox_ineq} imposes a maximum allowed $t_{ox}$ for a given $t_{FE}$ (or vice-versa, a minimum $t_{FE}$ for a given $t_{ox}$), to achieve $MW>0$. This is visualized in \cref{fig:1}, that shows the transition between hysteretic regime (i.e., $MW>0$) and negative capacitance regime: \eqref{eq:tox_ineq} is in fact the opposite condition to that of stable negative-capacitance operation \cite{Hoffmann2018}. 
    
    Note that to arrive at the simple approximate closed-form expression in \eqref{eq:tox_ineq}, we assumed dominant inversion charge in the semiconductor that allowed considering the MOSFET capacitance to be equal to the oxide interlayer capacitance $C_{ox}$. In general however, the constraint as expressed in \eqref{eq:tox_ineq} is affected by the additional series capacitance provided by the semiconductor body of the underlying MOSFET \cite{Cao2020}, leading to a non-linear bias dependent $t_{ox}^*$. 
\begin{figure}[t!]
    \centering{
        \includegraphics{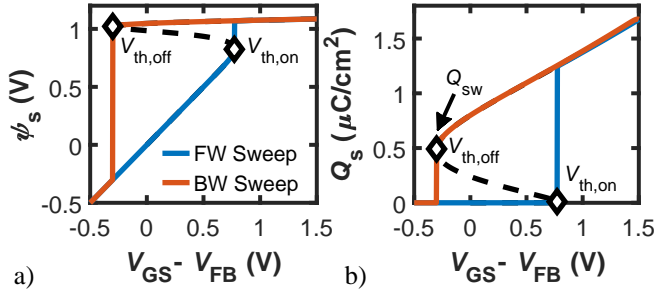}
        \subfloat{\label{fig:r2_a}}
        \subfloat{\label{fig:r2_b}}
    }
    \caption{\protect\subref{fig:r2_a} $\psi_s$ and \protect\subref{fig:r2_b} $Q_s$ vs $V_{GS}-V_{FB}$, showing the non-symmetric switching conditions at $V_{th,on}$ and $V_{th,off}$ (the simulation parameters are reported in \cref{tab:sim_param}, $t_{FE}=\SI{10}{\nano\metre}$). The black dashed line between the forward (FW) and backward (BW) branches shows the negative capacitance region, which is an unstable region of operation, see \eqref{eq:tox_ineq}.}
    \label{fig:r2}
\end{figure}
\begin{figure}[t!]
    \centering{
        \includegraphics[width=\columnwidth]{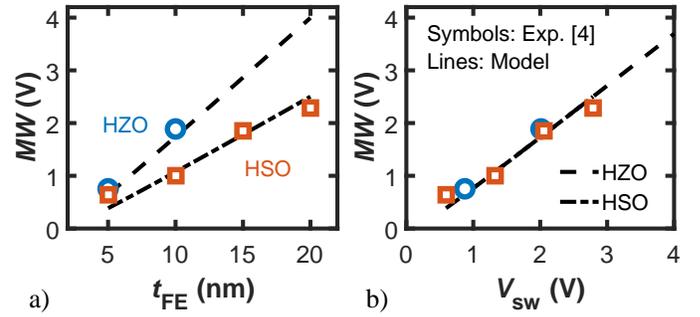}
    }
    \subfloat{\label{fig:r3_a}}
    \subfloat{\label{fig:r3_b}}
    \caption{Comparison of $MW$ calculated with \eqref{eq:mw} and $MW$ experimental data from \cite{Ali2018} plotted against \protect\subref{fig:r3_a} $t_{FE}$ and \protect\subref{fig:r3_b} switching voltage, $V_{sw}$ (defined in \eqref{eq:vsw}), revealing an universal $MW$ scaling behavior regardless of the particular ferroelectric material.}
    \label{fig:r3}
\end{figure}
\subsection{Applicability Limits of the Modeling Approach}\label{sec:anal_validity}
As specified in \cref{sec:introduction}, the derivation of the analytical expressions \eqref{eq:vth_on}-\eqref{eq:vth_off}, \eqref{eq:mw} was carried out starting from the Landau-Devonshire phenomenological theory (also known as ‘single-domain’ approximation) which treats ferroelectric as an homogeneous layer where, under the specific conditions discussed in \cref{sec:anal_derivation}, switching occurs between the two stable saturated polarization values \cite{Salahuddin2008}. In general, non-uniform polarization present in realistic ferroelectric thin layers can only be captured with the generalized Landau-Ginzburg theory that includes a domain interaction term in the expression of the free-energy \cite{Hoffmann2018}. However, recent attempts in the literature such as \cite{Gomez2019} demonstrate that it is possible to equivalently reproduce the effect of multi-domain interaction (which leads to gradual polarization switching) with multiple parallel single-domain models by considering finite ferroelectric switching time. In this work, we restricted the analysis to 'empirically' reproduce $MW$ of realistic FeFETs with effective $\alpha$, $\beta$ parameters that are able to capture the switching behavior of the saturated loops and neglecting the non-idealities that could reduce $MW$ (i.e., counteracting trapping phenomena, wake-up of ferroelectric and other effects \cite{Ali2018}).

Another important aspect related to ferroelectric HfO\textsubscript{2} is the polycrystalline (i.e., amorphous) structure of realistic layers, which leads to fluctuations in properties of ferroelectric (such as the coercive field, $E_C$) \cite{Chatterjee2019}. Although beyond the scope of this work, we mention that the analytical model can be used to investigate the effect of $E_C$ variations, for example, by carrying out the derivation of \eqref{eq:mw} with respect to both $\alpha$, $\beta$ considering that $E_C\approx-4/3\alpha\sqrt{-\alpha/6\beta}$ with $\varepsilon_{FE}=0$ \cite{Pahwa2017}.
\begin{figure*}[t!]
    \centering{
        \includegraphics{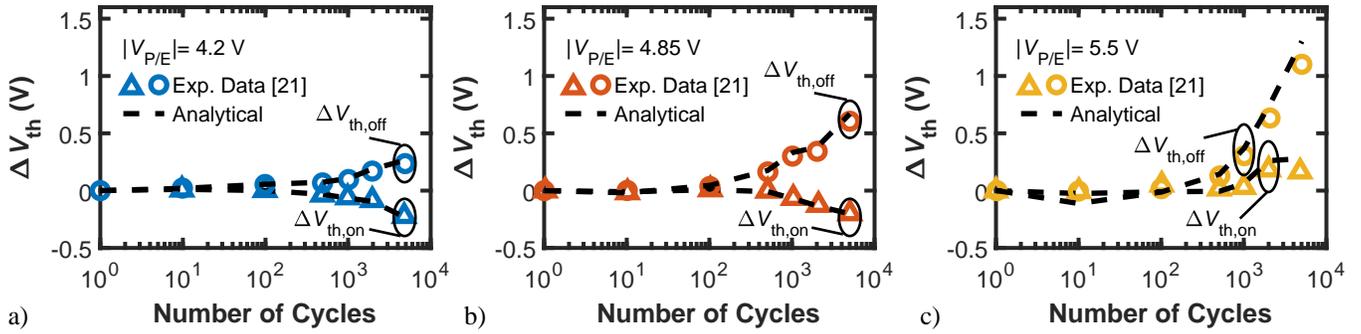}
    }
    \subfloat{\label{fig:r4_a}}
    \subfloat{\label{fig:r4_b}}
    \subfloat{\label{fig:r4_c}}
    \caption{Calculated (dashed lines) and measured (symbols) $\Delta{}V_{th,on}$ and $\Delta{}V_{th,off}$ vs program/erase cycle number. Experimental data is taken from \cite{Zeng2019}. The different panels show different program/erase pulse amplitude: \protect\subref{fig:r4_a} $|V_{P/E}|= \SI{4.2}{\volt}$, \protect\subref{fig:r4_b} $|V_{P/E}|= \SI{4.85}{\volt}$, and \protect\subref{fig:r4_c} $|V_{P/E}|= \SI{5.5}{\volt}$, respectively.}
    \label{fig:r4}
\end{figure*}
\subsection{Non-Symmetric Switching Conditions}\label{sec:anal_non_symmetric}
The lack of symmetry between $V_{th,on}$ and $V_{th,off}$ expressions is essentially caused by the non-linear $\psi_s$ and $Q_s$ vs $V_{GS}$ curves, see \cref{fig:r2}. At $V_{th,off}$, see \cref{fig:r2}\subref{fig:r2_b}, $Q_s$ is equal to the critical $Q_{sw}$ value and this leads to a linear dependence of $V_{th,off}$ on $V_{sw}$ as expressed by \eqref{eq:vth_off}. Conversely, since $Q_s(V_{th,on})>0$, $V_{th,on}$ cannot be proportional to $V_{sw}$ because this would require $Q_s<0$ (if $Q_{sw}$ is positive then $V_{sw}$ is negative, and vice-versa). $Q_s<0$ could only occur in the accumulation region, for $V_{GS}<V_{FB}$. In \cref{sec:anal_mw_max}, we discuss possible strategies to maximize $MW$ based on the optimization of switching conditions.
\subsection{Guidelines to Maximize MW}\label{sec:anal_mw_max}
The most obvious design strategy to increase $MW$ is to increase $t_{FE}$, as reported in \cite{Mulaosmanovic2019}, where a \SI{20}{\nano\metre}-thick ferroelectric was employed to roughly double $MW$. However, this solution goes in contrast with the need of scaling the technology. With the aid of the analytical expressions derived in \cref{sec:anal_derivation} it is possible to devise $MW$ maximization strategies without compromising FeFET scaling.

As mentioned previously, the first strategy is based on the consideration that non-symmetric switching conditions reduce the maximum $MW$ because $V_{th,on}$ is not proportional to $V_{sw}$. As also pointed out in \cite{Chen2011}, the theory predicts that another hysteresis loop can form between accumulation and depletion region that is basically symmetrical to the one from inversion to depletion (as the accumulation charge also depends exponentially on $\psi_s$). Thus, in principle, if a FeFET could switch from accumulation to inversion (and vice-versa), skipping the depletion region, then the switching conditions would become symmetrical and $MW$ would consequently increase. To achieve this, the condition $V_{th,on}<V_{FB}$ [with $V_{th,on}$ defined as in \eqref{eq:vth_on}] would have to be satisfied.

Another possible way to increase $MW$ at the same $t_{FE}$ and approaching the maximum theoretical limit \cite{Lue2002,Sallese2004}:
\begin{equation}\label{eq:mw_max}
    MW_{MAX}=2E_C\times{}t_{FE},    
\end{equation}
is to engineer the insulator layer between the ferroelectric and the semiconductor. This goal can be achieved by either scaling $t_{ox}$ or increasing $\varepsilon_{ox}$ (i.e., by employing high-$\kappa$ insulators). In the limit, the oxide layer should be removed to maximize $MW$; in fact, with $t_{ox}\rightarrow{}0$ then $MW$ would increase of $\approx50\%$ for $t_{FE}=\SI{10}{\nano\metre}$ and the ferroelectric parameters reported in \cref{tab:sim_param}.
\section{Results}\label{sec:results}
\subsection{Geometrical and Universal Scaling of MW}\label{sec:scaling_trends}
To verify the scaling trends of $MW$ vs $t_{FE}$ predicted by \eqref{eq:mw} we compared the analytical results with experimental data recently published in \cite{Ali2018} of FeFETs realized with Zr- and Si- doped HfO\textsubscript{2} (i.e., HZO and HSO) ferroelectrics. \cref{fig:r3}\subref{fig:r3_a} shows the experimental $MW$ vs $t_{FE}$ data points (symbols) taken from \cite{Ali2018} and the results obtained from \eqref{eq:mw} (lines).
The ferroelectric parameters were set as follows: $\alpha_{HZO}=\SI{-3e9}{\metre/\farad}$, $\beta_{HZO}=\SI{5e11}{\metre^5\per\farad\per\coulomb\squared}$ and $\alpha_{HSO}=\SI{-3.1e9}{\metre/\farad}$, $\beta_{HSO}=\SI{1.7e12}{\metre^5\per\farad\per\coulomb\squared}$, respectively. In both cases, $\varepsilon_{FE}$ was set to 16 \cite{Kao2018}. With these sets of $\alpha,\,\beta$ the calculated remnant polarization $P_r\approx{}\sqrt{-\alpha/2\beta}$ \cite{Pahwa2017} is in the range of $3-5\si{\micro\coulomb/\centi\metre\squared}$. These values are lower than the $P_r$ normally extracted for a MFM capacitor \cite{Ali2018}. This discrepancy might be due to the fact that the ferroelectric in an MFIS structure normally operates in a $P-E$ subloop and that the polarization is lower than the maximum achievable by the ferroelectric itself (due to the lower field on the ferroelectric in the FeFET) \cite{Ni2018}.

Since $V_{sw}\propto{}t_{FE}$ \eqref{eq:vsw}, and $MW\sim{}V_{sw}\propto{}t_{FE}$, our model correctly anticipates the experimentally observed linear thickness-dependence of the $MW$. Equation \eqref{eq:mw} also suggests that regardless of the material and geometrical parameters, the $MW$ should be only function of $V_{sw}$ (on a first order approximation). \cref{fig:r3}\subref{fig:r3_b} indeed reveals the universal trend of $MW$ vs $V_{sw}$ for different FE films. This important result comes from the fact that $V_{sw}$ embeds the specific ferroelectric and geometric parameters and implies that regardless of the technology the scaling follows the same trend.
\subsection{Assessing Endurance from MW Expression}\label{sec:endurance_assessment}
\begin{figure}[t!]
    \centering{
        \includegraphics{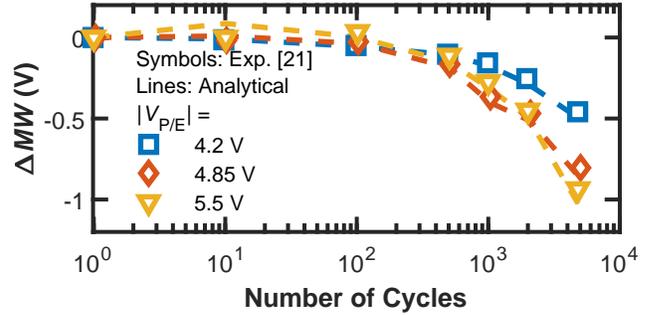}
    }
    \caption{Calculated (dashed lines) and measured (symbols) $\Delta{}MW$ vs program/erase cycle number. Experimental data is taken from \cite{Zeng2019}.}
    \label{fig:r5}
\end{figure}
As discussed in the Introduction, commercialization of FeFET has been hindered by the limited retention and endurance with respect to other technologies. Retention is defined as the time taken for the different states to be no longer distinguishable during a prolonged read operation. Instead, endurance is the time taken before states are indistinguishable after repeated program/erase operations. HfO\textsubscript{2}-based FeFETs have reduced trapping and lower depolarization over coercive field ratio with respect to PZT- or SBT-based devices leading to improved retention time \cite{Gong2016}. However, endurance is still a major issue for this technology imposing an upper limit of $\sim10^4-10^6$ writing cycles \cite{Yurchuk_TED_2014, Mulaosmanovic2019} that is far from meeting the International Roadmap for Devices and Systems (IRDS) requirements of $10^{12}$ cycles \cite{IEEE2018}. At the basis of the limited endurance lies the increasing trapping due to the generation of oxide and interface states in the layer between the ferroelectric and the semiconductor body \cite{Yurchuk_IRPS_2014}. This is a consequence of the lower dielectric constant of SiO\textsubscript{2} compared to that of doped-HfO\textsubscript{2}, that causes the local electric field to increase, accelerating generation of defects.

In the following, we derive an expression for the degraded $MW$ during endurance tests. Fast $MW$ decay due to depolarization fields and trapping/detrapping was not explicitly included in the model as it is expected to mainly influence retention rather than endurance \cite{Yurchuk_IRPS_2014,Gong2016}. Our analysis focuses on both oxide and interface traps generation during these tests, in which the gate voltage is cycled with program and erase pulses to induce ferroelectric switching. The prolonged effect of high voltage pulses over time induces degradation in the form of generation of defects, and this is modeled with the analytical formula derived in \cref{sec:anal_model} by adding the contribution due to the defects in the right-hand side of the SPE \eqref{eq:spe} \cite{Esqueda2014}:
\begin{equation}\label{eq:vdt}
    V_{ot}\equiv{}-\frac{q\Delta{N_{ot}}}{C_{ox}}\quad{}V_{it}\equiv{}\frac{q\Delta{}D_{it}}{C_{ox}}(\psi_s-\phi_b)
\end{equation}
where $\Delta{}N_{ot}$ is the generated trap concentration in the oxide interface layer (\si{\per\centi\metre\squared}), $\Delta{}D_{it}$ is the generated interface trap density of states (\si{\per\centi\metre\squared\per\electronvolt}), and $\phi_b$ is the body potential. These expressions assume that the charge neutrality level for the interface traps is located at Si mid-gap \cite{Esqueda2014}. The stress causing generation of traps is induced by positive and negative pulses on the gate performing erase and program operations in the FeFET, respectively. Hence, $V_{th,on}$ will tend to decrease and $V_{th,off}$ to increase \cite{Zeng2019}. The concentration of generated defects during writing of the memory is in general different depending on the sign of the writing pulse, therefore the shifts in $V_{th,on}$ and $V_{th,off}$ will not be symmetric. Thus, we will use different symbols to indicate the generated defects during program and erase cycles, namely $\Delta{}N_{ot,P\text{/}E}$ and $\Delta{}D_{it,P\text{/}E}$ for oxide and interface traps, respectively.

The degraded $V_{th,on}$, $V_{th_off}$, and $MW$ expressions are derived by rewriting the threshold conditions, taking into account the additional potential drop due to defects expressed in \eqref{eq:vdt} (the derivation is omitted for brevity). The $V_{th}$'s and $MW$ degradation is expressed as follows:
\begin{subequations}\label{eq:degraded}
    \begin{align}
        \Delta{}V_{th,on}=&2V_t\ln{\left(1+\frac{q\Delta{}D_{it,P}}{C_{ox}}\right)}\times\left(1+\frac{q\Delta{}D_{it,P}}{C_{ox}}\right)\notag\\
		-\frac{q}{C_{ox}}\times&\left\lbrace{}\Delta{}N_{ot,P}-\Delta{}D_{it,P}\left[2V_{t}\ln{\left(\frac{2V_{t}}{|a|Q_0}\right)}-2V_t-\phi_b\right]\right\rbrace{}\label{eq:vth_on_prime}\\
		\Delta{}V_{th,off}=&-\frac{q}{C_{ox}}\left\lbrace{}\Delta{}N_{ot,E}-\Delta{}D_{it,E}\left[2V_{t}\ln{\left(\frac{Q_{sw}}{Q_0}\right)}-\phi_b\right]\right\rbrace\label{eq:vth_off_prime}\\
		\Delta{}MW=&2V_t\ln{\left(1+\frac{q\Delta{}D_{it,P}}{C_{ox}}\right)}\times\left(1+\frac{q\Delta{}D_{it,P}}{C_{ox}}\right)\notag\\
		-\frac{q}{C_{ox}}&\left\lbrace(\Delta{}N_{ot,P}-\Delta{}N_{ot,E})-2V_{t}\Delta{}D_{it,P}\left[\ln{\left(\frac{2V_{t}}{|a|Q_0}\right)}-1\right]\right.\notag\\
		&\left.+2V_{t}\Delta{}D_{it,E}\ln{\left(\frac{Q_{sw}}{Q_0}\right)}+(\Delta{}D_{it,P}-\Delta{}D_{it,E})\phi_b\right\rbrace.\label{eq:mw_prime}
    \end{align}    
\end{subequations}
To assess the accuracy of the above expressions, we compared the analytical results with experimental data of endurance tests from \cite{Zeng2019}. The results in terms of $\Delta{}V_{th,on}$ and $\Delta{}V_{th,off}$ for three different values of program/erase pulse amplitude, $|V_{P/E}|$ are shown in \cref{fig:r4} [$|V_{P/E}|=$ \SI{4.2}{\volt} \subref{fig:r4_a}, \SI{4.85}{\volt} \subref{fig:r4_b}, and \SI{5.5}{\volt} \subref{fig:r4_c}]. The $\alpha$, $\beta$ values set to match the experimental data trends are \SI{-2.3e9}{\meter/\farad} and \SI{1e12}{\meter^5/\farad/\coulomb^2}, respectively. The duration of both program and erase pulse for each $|V_{P/E}|$ is $t_{P/E}=\SI{100}{\nano\second}$, thus the time it takes for a single writing cycle is $t_{cycle}=\SI{200}{\nano\second}$ \cite{Zeng2019}. The combination of $V_{th,on}$ and $V_{th,off}$ degradation affects $MW$ in turn, as shown in \cref{fig:r5} for the same $|V_{P/E}|$ values of \cref{fig:r4}.

The trend of the degraded thresholds and $MW$ is fully captured by $V_{ot}$ and $V_{it}$ only. This happens because degradation primarily occurs in the insulator layer, as discussed previously. From this observation, simplified formula can be derived for $\Delta{}V_{th,on}$, $\Delta{}V_{th,off}$ and $\Delta{}MW$. By neglecting $V_{ins}$ variations in the modified SPE, \eqref{eq:vth_on_prime}-\eqref{eq:mw_prime} can be simplified as follows: 
\begin{subequations}\label{eq:degraded_simple}
	\begin{align}
		&\Delta{}V^\prime_{th,on}\sim\frac{-q}{C_{ox}}\left\lbrace{}\Delta{}N_{ot,P}-\Delta{}D_{it,P}\left[2V_{t}\ln{\left(\frac{2V_{t}}{|a|Q_0}\right)}-\phi_b\right]\right\rbrace\label{eq:dvth_on_approx}\\
		&\Delta{}V^\prime_{th,off}\sim\frac{-q}{C_{ox}}\left\lbrace{}\Delta{}N_{ot,E}-\Delta{}D_{it,E}\left[2V_{t}\ln{\left(\frac{Q_{sw}}{Q_0}\right)}-\phi_b\right]\right\rbrace\label{eq:dvth_off_approx}\\
		&\Delta{}MW^\prime\sim\frac{-q}{C_{ox}}\left\lbrace\left(\Delta{}N_{ot,P}-\Delta{}N_{ot,E}\right)\vphantom{\frac{2V_{t}}{aQ_0}}\right.-\Delta{}D_{it,P}\left[2V_{t}\ln{\left(\frac{2V_{t}}{|a|Q_0}\right)}\right]\notag\\
		&\left.\vphantom{\ln{\left(\frac{Q_{sw}}{Q_0}\right)}}+\Delta{}D_{it,E}\left[2V_t\ln{\left(\frac{Q_{sw}}{Q_0}\right)}\right]+\left(\Delta{}D_{it,P}-\Delta{}D_{it,E}\right)\phi_b\right\rbrace\label{eq:dmw_approx}.
	\end{align}
\end{subequations}
Note that $\Delta{}V^\prime_{th,on}$, $\Delta{}V^\prime_{th,off}$, and $\Delta{}MW^\prime$ are proportional to the variation introduced by the generation of both oxide and interface defects. The surface potential $\psi_s$ [corresponding to the logarithmic terms in square brackets in \eqref{eq:dvth_on_approx}-\eqref{eq:dvth_off_approx}] is calculated differently according to the two threshold conditions defined in \cref{sec:anal_derivation}. As intuition suggests, if the degradation were symmetric, i.e., the generated defects were giving equal and opposite in sign $V_{ot}$ and $V_{it}$, the $MW$ variation would be $\sim-2q/{C_{ox}}\times\left[\Delta{}N_{ot}-\Delta{}D_{it}\left(\psi_s-\phi_b\right)\right]$. 

The good agreement between analytical and experimental results in \cref{fig:r4,fig:r5} was obtained by extracting the generated oxide and interface trap concentrations from $\Delta{}V_{th,on}$ and $\Delta{}V_{th,off}$ data in \cite{Zeng2019} following the approach described in \cite{Schroder2005}. 
That is, $N_{ot}$ and $D_{it}$ were extracted by separating the threshold voltage shifts due to oxide ($\Delta{}V_{mg}$) and interface traps ($\Delta{}V_{it}$) separately. The former is obtained from the mid-gap voltage, $V_{mg}$, that correlates with $N_{ot}$-induced $V_{th}$ drifts as at $V_G=V_{mg}\Rightarrow\psi_s=\phi_b$ and $\Delta{}V_{it}=0$, see \eqref{eq:vdt}; the latter is obtained by $\Delta{}V_{it}=\Delta{}V_{th}-\Delta{}V_{ot}$ \cite{Zeng2019,Schroder2005}. To summarize, \eqref{eq:vth_on_prime}-\eqref{eq:mw_prime} transparently connect the FeFET parameters to the stress-dependent oxide and interface trap generation. As such, \eqref{eq:mw_prime} offers a powerful new $MW$-based characterization tool for extracting oxide and interface defects. This could serve either as an alternative to traditional techniques, or a stand-alone method to characterize defect densities under a variety of stress conditions. For instance, notice that when only $N_{ot}$ generation affects $MW$ degradation then it is possible to estimate the \emph{net} generated traps from \eqref{eq:dmw_approx}:
\begin{equation}\label{eq:d_not}
    \Delta{}N_{ot,net}\equiv{}\Delta{}N_{ot,P}-\Delta{}N_{ot,E}\approx-\Delta{}MW^\prime{}\frac{C_{ox}}{q}
\end{equation}
This expression allows to simply and directly correlate $MW$ measurements with generated traps. In the next section, we will exploit \eqref{eq:d_not} to provide endurance predictions.
\subsection{Writing Conditions Agnostic Endurance}\label{sec:endurance_prediction}
\begin{figure}[t!]
    \centering{
        \includegraphics{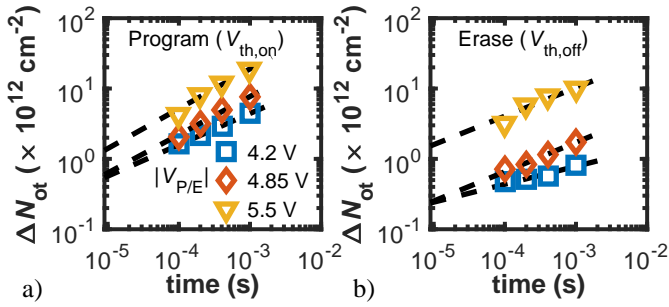}
    }
    \subfloat{\label{fig:r6_a}}
    \subfloat{\label{fig:r6_b}}
    \caption{Generated oxide traps, $\Delta{}N_{ot}$, vs program/erase time for different $|V_{P/E}|$ (see legend) determining \protect\subref{fig:r6_a} $V_{th,on}$ and \protect\subref{fig:r6_b} $V_{th,off}$ degradation. Black dashed lines are the fitting of experimental data (symbols, from \cite{Zeng2019}) with \eqref{eq:not_pow_law}.}
    \label{fig:r6}
\end{figure}
\begin{figure}[t!]
    \centering{
        \includegraphics{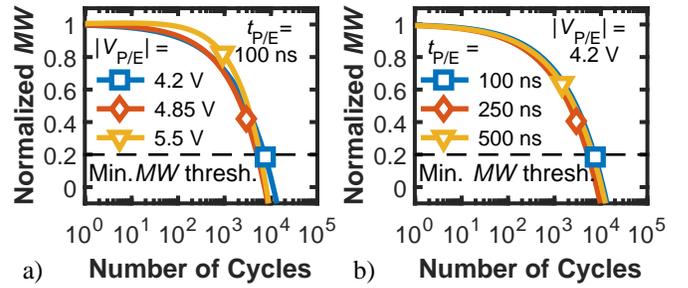}
    }
    \subfloat{\label{fig:r7_a}}
    \subfloat{\label{fig:r7_b}}
    \caption{Normalized $MW$ degradation calculated from \eqref{eq:dmw_approx} with only the contribution of $\Delta{}N_{ot}$ extrapolated from \cref{fig:r6}. \protect\subref{fig:r7_a} and \protect\subref{fig:r7_b} show the dependence for different $|V_{P/E}|$ and $t_{P/E}$ values, respectively. An arbitray minimum $MW$ threshold is identified to define endurance.}
    \label{fig:r7}
\end{figure}

In the following we will show that the endurance extrapolated from the equations derived previously is not influenced by the writing conditions (in terms of $|V_{P/E}|$ and $t_{P/E}$). With the $N_{ot}$ and $D_{it}$ data extracted in \cref{sec:endurance_assessment}, it is possible to extrapolate the generated trap concentration for an arbitrary number of writing cycles. For simplicity and clarity of presentation, we will assume that the $MW$ degradation is induced by oxide traps only (as supported by the experimental data in \cite{Zeng2019}) and neglect the generation of interface traps. The generated oxide trap density, $N_{ot}$ is shown in \cref{fig:r6}\subref{fig:r6_a}, \subref{fig:r6_b} for both program and erase operation that set $V_{th,on}$ and $V_{th,off}$, respectively. 
By fitting the experimental data in \cref{fig:r6} it is found that generated oxide trap concentration follows a power law with respect to writing time (the duration of a single writing cycle being $t_{cycle}=\SI{200}{\nano\second}$ \cite{Zeng2019}):
\begin{equation}\label{eq:not_pow_law}
    \Delta{}N_{ot}\sim{}N_0\times{}(t_{cycle})^{\beta_s}
\end{equation}
where $N_0$ and $\beta_s$ are coefficients to fit experimental data, whose values for different writing conditions are collected in \cref{tab:power_law}. Exponent $\beta_s$ in the range $0.3-0.5$ might be a signature of enhanced TDDB due to repeated cycling as reported in \cite{Kerber2010}. The extrapolated $MW$ degradation obtained by using the predicted $\Delta{}N_{ot}$ from the generation model is shown in \cref{fig:r7}\subref{fig:r7_a}, \subref{fig:r7_b} for different $V_{P/E}$ and $t_{P/E}$ values, respectively. 
\begin{table}[t!]
\caption{Coefficients of the power law in \cref{eq:not_pow_law}.}
\label{tab:power_law}
\begin{center}
\begin{tabular}{@{}ccccccc@{}}
\toprule
& \phantom{abc} & \multicolumn{2}{c}{\textbf{Program (\emph{V\textsubscript{th,on}})}} & \phantom{abc} & \multicolumn{2}{c}{\textbf{Erase (\emph{V\textsubscript{th,off}})}}\\
\cmidrule{3-4} \cmidrule{6-7}
\multicolumn{2}{c}{$V_{P/E}$ (\si{\volt})} & $N_0$ (\si{\centi\metre^{-2}}) & $\beta_s$ & & $N_0$ (\si{\centi\metre^{-2}}) & $\beta_s$\\
\midrule
\multicolumn{2}{c}{4.2} & \SI{9.6e13}{} & 0.45 & & \SI{4.6e12}{} & 0.25 \\
\multicolumn{2}{c}{4.85} & \SI{3.28e14}{} & 0.54 & & \SI{3.1e13}{} & 0.41 \\
\multicolumn{2}{c}{5.5} & \SI{9.5e14}{} & 0.54 & & \SI{1.7e14}{} & 0.41 \\
\bottomrule
\end{tabular}
\end{center}
\end{table}
Note that $MW$ values are normalized to the respective initial value for a fair comparison with different writing conditions. The FeFET is considered to fail to retain its memory operation after reaching the arbitrary minimum $MW$ threshold set as the 20\% of the initial value, see \cref{fig:r7}. Interestingly, from \cref{fig:r7}\subref{fig:r7_a} it appears that $|V_{P/E}|$ increase does not degrade endurance significantly (at least for the range of values as in \cite{Zeng2019}). This is because higher $|V_{P/E}|$ leads to higher initial $MW$ \cite{Ali2018} but also higher $\Delta{}N_{ot}$, see \cref{fig:r7}. Similarly, \cref{fig:r7}\subref{fig:r7_b} shows that increasing the pulse duration negligibly influences endurance. Note that in this case it was assumed that $t_{P/E}$ increase leads to the same increase in $MW$ and initial $N_{ot}$ to that caused by $V_{P/E}$. This was done for the specific purpose of illustrating that if both $MW$ and initial $N_{ot}$ increase with program conditions, then the combined effect leads to negligible variation in endurance. Note that in general, if the assumption regarding $MW$ and $N_{ot}$ increase with $V_{P/E}$ (or $t_{P/E}$) is not satisfied, then the endurance limit will be affected by the writing conditions.
The model can also predict the endurance improvement that can be obtained if the generated trap are decreased, either by improving the SiO\textsubscript{2}/Si interface quality or by reducing the field in the oxide layer. For example, if $N_0$ is decreased by one order of magnitude (and assuming every other parameter constant) then endurance can be extended to $10^6$ cycles. These considerations can be helpful to develop next generation FeFET with extended endurance.
\section{Conclusions}\label{sec:conclusions}
In this paper, we derived an analytical expression of the Memory Window, $MW$, that can be used to investigate the scaling trends and endurance limits of FeFETs. Based on the Landau-Devonshire formalism, we arrived at closed-form expressions for the threshold voltages, $V_{th,on/off}$, and $MW$ for a conventional Metal-Ferroelectric-Insulator-Semiconductor (MFIS) structure that depends on critical technological and geometrical parameters. The $MW$ expression also includes the effect of generated interface and oxide traps to assess the endurance limits of FeFETs. The key findings of this work are as follows:
\begin{enumerate}
    \item $MW$ as expressed in \eqref{eq:mw} is a material-independent universal function of switching voltage, $V_{sw}$, embedding the dependence on critical design parameters.
    \item Constraints on minimum ferroelectric thickness ($t_{FE}$) for a given oxide interface layer thickness ($t_{ox}$), see \eqref{eq:tox_ineq}, impose a trade-off between scaling and $MW$ amplitude.
    \item The $MW$ being lower than the theoretical limit expressed in \eqref{eq:mw_max} is due to the non-symmetrical switching conditions.
    \item From the $V_{th,on/off}$ analytical expressions in \eqref{eq:vth_on}-\eqref{eq:vth_off}, 	guidelines can be devised for $MW$ maximization (for a given $t_{FE}$) by engineering the non-symmetric switching conditions or the oxide interface layer.    
    \item $MW$ can be used to extract oxide and interface trap concentration that are generated during endurance tests, see \eqref{eq:d_not}.
    \item The generated traps increase as a power-law, see \eqref{eq:not_pow_law}, with time exponent $\sim0.3-0.5$. Under specific assumptions, the endurance limit is essentially independent of writing conditions.
\end{enumerate}

\appendices
\section{Validation of the Analytical Model}\label{sec:appendix_validation}
\begin{figure}[t!]
    \centering{
        \includegraphics{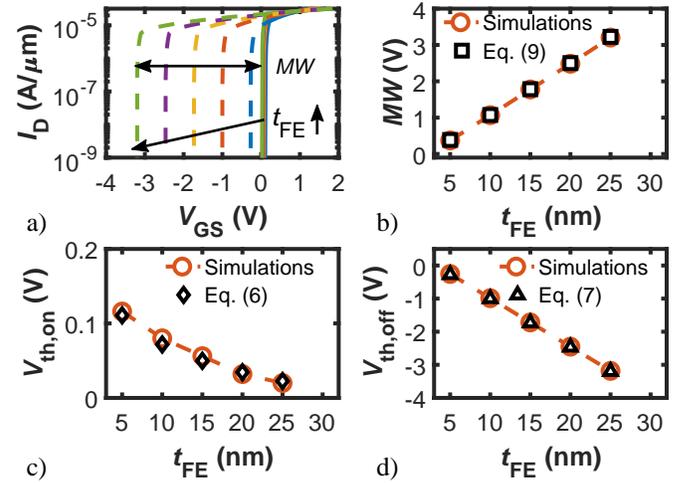}
    }
    \subfloat{\label{fig:r8_a}}
    \subfloat{\label{fig:r8_b}}
    \subfloat{\label{fig:r8_c}}
    \subfloat{\label{fig:r8_d}}
    \caption{\protect\subref{fig:r8_a} Simulated $I_D$ vs $V_{GS}$ curves for different $t_{FE}$. \protect\subref{fig:r8_b}, \protect\subref{fig:r8_c}, and \protect\subref{fig:r8_d} show the comparison between the $MW$, $V_{th,on}$, and $V_{th,off}$ obtained from simulations (orange circles) and from \cref{eq:mw}, \cref{eq:vth_on}, and \cref{eq:vth_off} (black squares, diamonds, and triangles), respectively.}
    \label{fig:r8}
\end{figure}

To verify the accuracy of the derived expression and of the underlying assumptions, we compared the analytical result of \eqref{eq:vth_on}-\eqref{eq:vth_off}, \eqref{eq:mw} with numerical simulations. The comparison was done with simulations based on the Landau-Devonshire theory described in \cref{sec:anal_model} to quantify the discrepancy with the analytical results. The simulations compute the self-consistent solution for $\psi_s$ from \eqref{eq:spe} coupled with the $Q_s$ expression \cite{Taur2009}. The Landau parameters are those used in \cref{sec:anal_derivation} for the Si-doped HfO\textsubscript{2} (i.e., HSO) (the full parameter set is collected in \cref{tab:sim_param}). The solution for $\psi_s$ and $Q_s$ was then used to calculate the drain current via the Pao-Sah double integral \cite{Chen2011,Taur2009}, see \cref{fig:r8}\subref{fig:r8_a} from which the trend of $MW$ with $t_{FE}$ was extracted, as shown in \cref{fig:r8}\subref{fig:r8_b}. Note that $t_{ox}=\SI{1}{\nano\metre}$ was chosen small enough to ensure hysteresis for the whole $t_{FE}$ range considered (as discussed in \cref{sec:anal_derivation}). The agreement between the simulations and the analytical expressions, see \cref{fig:r8}, shows that the approximations made in the derivation of \eqref{eq:vth_on}-\eqref{eq:mw} are acceptable. Remarkably, the analytical expressions predict a weak decrease of $V_{th,on}$ with increasing $t_{FE}$, whereas $V_{th,off}$ decreases linearly, see \cref{fig:r8}\subref{fig:r8_c}, \subref{fig:r8_d}. This behavior follows from the consideration made in \cref{sec:anal_non_symmetric} on the non-symmetric switching, regarding the different conditions under which $V_{th,on}$ and $V_{th,off}$ are derived.
\section{Comparison with Preisach Model}\label{sec:appendix_preisach}
\begin{figure}[t!]
    \centering{
        \includegraphics{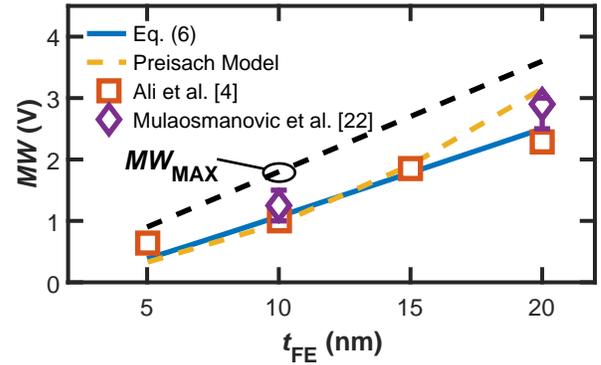}
    }
    \caption{Comparison between $MW$ obtained with \eqref{eq:mw} (blue solid line), Preisach Model (numerical simulations, yellow dashed line) and experimental data \cite{Ali2018,Mulaosmanovic2019}. The black dashed line is the theoretical maximum value, $MW_{MAX}$ (with $E_C=\SI{0.9}{\mega\volt/\centi\metre}$, \cite{Ali2018}), see \eqref{eq:mw_max}.}
    \label{fig:r9}
\end{figure}

Ferroelectric switching behavior is described in the literature also by other models; here we focus on the Preisach model \cite{Alam2019} which is broadly employed to interpret experimental results of FeFETs. 
In the framework of this model, the fact that measured $MW$ is lower than the theoretical limit, see \eqref{eq:mw_max}, is attributed to sub-hysteresis trajectories in the $P-V$ loop followed depending on the writing conditions \cite{Ni2018,Lue2002,Jiang1997}. The theoretical limit is thus not reached due to switching events with $E<E_C$. In the case of the Landau formalism followed in this work, the same result of $MW$ being lower than $MW_{MAX}$ can be ascribed to non-symmetric switching conditions, as explained also in \cref{sec:anal_non_symmetric}. This is supported by the comparison of the analytical $MW$ results with numerical simulations carried out with a commercial software \cite{Synopsys2018} on an MFIS structure with the Preisach model (the same parameter set was used, see \cref{tab:sim_param}). We performed numerical simulations with the Preisach model because, unlike with \eqref{eq:mw}, it is not possible to derive a closed-form solution for the $MW$, because the switching points for the inner loops depend on the ferroelectric history and cannot be determined \emph{a priori}. The comparison between the analytical expression derived in this work, the Preisach model and experimental data (from \cite{Ali2018,Mulaosmanovic2019}) shown in \cref{fig:r9} confirms the fact that $MW$ obtained both with Preisach and Landau model is below the theoretical limit $MW_{MAX}$.
\begin{table}[t!]
\caption{Parameters used in the numerical simulations of \cref{fig:r8}.}
\label{tab:sim_param}
\begin{center}
\begin{tabular}{@{}lcccr@{}}
\toprule
\textbf{Symbol} & \phantom{abc} & \phantom{abc} & \phantom{abc} &\textbf{Value}\\
\midrule
$L_G,\; W_G$ (\si{\micro\metre}) & & & &\num{1} \\
$N_a$ (\si{\per\cubic\centi\metre}) & & & & \num{5e17} \\
$\mu_n\;(\si{\centi\metre\squared/\volt\second})$ & & & & \num{200}\\
$V_{FB}\;(\si{\volt})$ & & & & \num{-0.7} \\
$t_{ox}\;(\si{\nano\metre})$ & & & & 1 \\
$t_{FE}\;(\si{\nano\metre})$ & & & & $(5\var{-}25)$ \\
$\alpha\;(\si{\metre/\farad})$ & & & & \num{-3.1e9} \\
$\beta\;\left(\si{\metre^5/\farad/\coulomb\squared}\right)$ & & & & \num{1.7e12}\\
$\varepsilon_{FE}\;(1)$ & & & & \num{16} \cite{Kao2018}\\
\bottomrule
\end{tabular}
\end{center}
\end{table}
\section{Outline of the derivation of \eqref{eq:vth_on_prime}-\eqref{eq:vth_off_prime}}\label{sec:appendix_degr_mw_deriv}
As mentioned in \cref{sec:endurance_assessment}, the expressions for $\Delta{V_{th,on}}$, $\Delta{V_{th,off}}$, see \eqref{eq:vth_on_prime}-\eqref{eq:vth_off_prime}, were derived by following the same procedure of \cref{sec:anal_derivation} by modifying \eqref{eq:spe} as follows:
\begin{equation}\label{eq:spe_degr}
    V_{GS}-V_{FB}=V_{ins}-\frac{q\Delta{N_{ot,P/E}}}{C_{ox}}+\frac{q\Delta{D_{it,P/E}}}{C_{ox}}(\psi_s-\phi_b)
\end{equation}
with the same symbols as previously defined. For $\Delta{}V_{th,on}$, the expression for the charge at the switching condition ($\diffp{V_{GS}}/{\psi_s}=0$) reads:
\begin{equation}\label{eq:q_vthon_degr}
    Q_s(\Delta{}V_{th,on})=-\frac{2V_t}{a}\left(1+\frac{q\Delta{}D_{it,P}}{C_{ox}}\right).
\end{equation}
By substituting \eqref{eq:q_vthon_degr} and the corresponding $\psi_s$ (obtained as $\psi_s=2V_t\ln{(Q_s/Q_0)}$) in \eqref{eq:spe_degr}, one obtaines \eqref{eq:vth_on_prime}. For $\Delta{}V_{th,off}$, the switching condition ($\diffp{Q_s}/{V_{ins}}=0$) does not alter $Q_{sw}$ and corresponding $\psi_s$ expressions. Thus, \eqref{eq:vth_off_prime} is simply obtained by substituting $\psi_s=2V_t\ln{Q_{sw}/Q_0}$ in \eqref{eq:spe_degr}. The expression for $\Delta{}MW$, see \eqref{eq:mw_prime}, is obtained by subtracting \eqref{eq:vth_off_prime} from \eqref{eq:vth_on_prime}.
\section*{Acknowledgment}
The authors thank Thomas Mikolajick (NaMLab, TU Dresden), Francesco Maria Puglisi (University of Modena and Reggio Emilia), and Kamal Karda (Purdue University) for the valuable discussions.

\end{document}